\documentstyle[12pt,times]{article}

              \def\d{\delta}
   \def\e{\epsilon}        
\def\g{\gamma}   \def\G{\Gamma}   \def\k{\kappa}     \def\l{\lambda}
  \def\m{\mu}      \def\n{\nu}        \def\r{\varrho}
\def\o{\omega}      \def\p{\psi}       
       \def\t{\tau}
 \def\x{\xi}            \def\th{\theta}

\def\CA{{\cal A}}   
 \def\CF{{\cal F}}  \def\CG{{\cal G}}
\def\CH{{\cal H}}   \def\CL{{\cal L}}
\def\CN{{\cal N}}
\def\CO{{\cal O}}

\def\slash#1{\,/\kern-7pt#1}
\def\rd{\partial}

\def\darr#1{\raise1.5ex\hbox{$\leftrightarrow$}\mkern-16.5mu #1}

\def\rds{/\kern-6pt\rd}
\newcommand{\be}{\begin{equation}}
\newcommand{\bea}{\begin{eqnarray}}
\newcommand{\ee}{\end{equation}}
\newcommand{\eea}{\end{eqnarray}}
\newcommand{\nn}{\nonumber}

\newcommand{\id}{\kern0.2em\rule{0.1mm}{0.71em}
                 \kern0.12em\rule{0.1mm}{0.71em}
                 \kern-0.27em\rule[0.68em]{0.27em}{0.1mm}
                 \kern-0.30em\rule{0.44em}{0.1mm}\rule{0.1em}{-1mm}}

\makeatletter

\@addtoreset{equation}{section}
\makeatother                      

\begin{document}

\baselineskip 6mm
\renewcommand{\thefootnote}{\fnsymbol{footnote}}

\begin{titlepage}

\hfill\parbox{4cm}
{ KIAS-P99076 \\ hep-th/99mmnnn \\ August 1999}

\vspace{15mm}
\begin{center}
{\Large \bf The Supercharges of Eleven-dimensional Supergraviton 
on Gravitational Wave Background}
\end{center}

\vspace{5mm}
\begin{center} 
Seungjoon Hyun\footnote{\tt hyun@kias.re.kr}
\\[5mm]
{\it 
School of Physics, Korea Institute for Advanced Study,
Seoul 130-012, Korea
}
\end{center}
\thispagestyle{empty}

\vfill
\begin{center}
{\bf Abstract}
\end{center}
\noindent
We find the explicit expression of the supercharges of eleven dimensional 
supergraviton on the background geometry of gravitational waves 
in asymptotically light-like 
compactified spacetime. We perform the calculations order by order in the 
fermions
$\p$, while retaining all orders in bosonic degrees of freedom, and get the 
closed form up to $\p^5$ order. This should correspond to the supercharge of
the effective action of 
(0+1)-dimensional matrix quantum mechanics for, at least, $v^4$ and $v^6$ 
order terms and their superpartners.
\vspace{2cm}
\end{titlepage}

\baselineskip 7mm
\renewcommand{\thefootnote}{\arabic{footnote}}
\setcounter{footnote}{0}

\section{Introduction}
One of the most striking observations in the recent developments 
in M/string theory is the realization of the deep connection between 
the supergravity and super Yang-Mills theory. 
Among others, one remarkable
example is the $AdS$/CFT correspondence\cite{maldacena}. 
It tells us 
that the M/String theory or their low energy effective supergravity on $AdS$ 
spaces are equivalent to the conformal
 limit of super Yang-Mills theory on the boundary of $AdS$ spaces. 
This holographic nature of 
 the theory has been clarified in \cite{susswit}. 

Another important example is matrix theory \cite{bfss} in which 
(0+1)-dimensional $SU(N)$ matrix quantum mechanics is conjectured 
to give eleven dimensional M/supergravity
in the large $N$ limit. In the framework of the discrete light-cone 
quantization (DLCQ), in which we take light-cone circle with radius $R$, the 
correspondence seems to hold even for finite $N$ \cite{susskind} and 
then one may understand the original matrix theory on full eleven-dimensional
non-compact spacetime as the large $N$ limit, 
while keeping the light-cone momentum $p_-=N/R$ fixed. 

 From the prescriptions given in 
\cite{seiberg, sen}, the DLCQ M theory on $T^p$ is described by 
$(p+1)$-dimensional super Yang-Mills theory on dual torus, 
for up to $p$=3. 
By applying their arguments on the supergravity backgrounds,
 it has been argued in \cite{hyun}
that, as a kind of generalization of $AdS$/CFT correspondence, these 
microscopic descriptions of DLCQ M theory on $T^p$ via super Yang-Mills 
theory corresponds to the M/string
theory on non-trivial backgrounds such as plane-fronted gravitational waves 
or $AdS$ spaces with some identifications.
This can be inferred from the simple observation that 
the limit considered in DLCQ M theory on
$T^p$  is exactly the same as the limit considered by Maldacena 
in \cite{maldacena} to get the $AdS$/CFT correspondence. It is the natural 
limit if one wants to have only
D-brane world-volume theory, while decoupling all the bulk degrees of freedom.
In \cite{maldacena}, the corresponding supergravity backgrounds are found by
near-horizon limit of the 
supergravity solutions in the presence of source branes. 
In \cite{hyun}, those come from  taking DLCQ and 
T-duality on the 
supergravity solutions of the source branes.

Especially interesting cases are DLCQ M theories on $T^0$ (noncompact 
ten-dimensional spacetime) and $T^1$, which 
correspond to the original matrix theory and matrix string 
theory \cite{DVV}, respectively.  
In these cases we
get the gravitational waves on the asymptotically light-like compactified 
spacetime as 
background geometry \cite{hyun,hks,bl,hk}. 
Indeed in \cite{bbpt}, it has been shown that the matrix quantum mechanics 
at one loop
gives the same effective action as those of the eleven-dimensional 
probe graviton on this background. Subsequently, superpartners of bosonic
$F^4$ term have been explicitly obtained within matrix theory and 
verified to agree
with those of supergraviton on this background 
\cite{harvey}-\cite{hks4}.

One may wonder why they give the correct descriptions, without good 
explanation on the nature of holography.   
Recent studies  on the non-renormalization theorem of matrix 
model due to 16 supercharges \cite{sethi2,sethi3,lowe} strongly suggests 
the key role played by the supersymmetry in 
these correspondences. After all, the regime that we can trust supergravity 
solutions is different from
the regime that we can trust Yang-Mills descriptions of D-branes, and 
only reasonable amount of supersymmetry would connect those two regions.

In this paper, we want to shed some light on these issues. In explicit, 
we find the supercharge of the eleven-dimensional supergraviton in the
background of gravitational waves, which are eleven-dimensionally lifted
D0 solutions. 
First of all, we consider the superparticle on flat background. 
The superparticle action we choose is the first quantized version of 
eleven-dimensional supergravity. If  the light-cone coordinate 
$x^-$  is periodically identified, the light-cone momentum is
quantized, $p_-={N \over R}$, and the single supergraviton ($N=1$)
moving in that direction is described by (0+1)-dimensional $U(1)$ 
matrix quantum mechanics.
The original superparticle action has 
target space supersymmetry, 
which is apparently very different from the supersymmetry of 
super Yang-Mills theory. However, the superparticle action 
has additional local world-line fermionic symmetry, namely $\k$-symmetry, 
and it is shown that they are identical 
after choosing the light-cone gauge for this local $\k$-symmetry and
 modifying the original supersymmetry by the appropriate $\k$-transformations
in such a way to preserve the gauge-fixing conditions.

In section 3 we consider  the superparticle 
on the background geometry
produced by source gravitational waves on the asymptotically light-like 
compactified spacetime. 
As a (0+1)-dimensional $\CN=16$ D0 quantum 
mechanics, this corresponds to the effective theory of probe D0-branes 
moving in the
background of $N$ source D0-branes. Since the full explicit 
expressions for the action
of the superparticle on non-trivial background are not known,
we perform the calculations order by order in fermions $\p$ and get 
supercharges up to $\psi^5$ order.  

Among others, we find the effective Lagrangian on this background is given by 
\bea
{1\over p_-}\CL_{eff}&=&{1\over 1+\sqrt{1-hv^2}}v^2
           +i{(1+\sqrt{1-hv^2})\over 4\sqrt{1-hv^2}}
           \psi\dot{\psi}+i{v^2 v_i\partial_j h(\psi\g^{ij}\psi)
           \over 8\sqrt{1-hv^2}(1+\sqrt{1-hv^2})}\nn\\
          &&-{h^2v^2\over 32(1-hv^2)^{3/2}}(\p\dot{\p})^2
           -{(2-hv^2)v_i\rd_j h\over32(1-hv^2)^{3/2}}
           (\psi\g^{ij}\psi)\p\dot{\p} +\cdots ~,
\label{eff}
\eea
where $x^i$ are the position coordinates of superparticle 
(Higgs fields in D0 quantum mechanics),  $v^i=\dot x^i$ and $h$ is the 
nine-dimensional harmonic function which will be given later. 

The Noether supercharges, quite surprisingly, turn out to be
\bea
Q= p\cdot\g\p +\CO(\p^5)~,
\label{charge1}
\eea
where $p_i\equiv {1\over p_-}{\rd \CL_{eff}\over \rd v^i}$ is 
the effective conjugate momentum of $x^i$ in units of $p_-$. 
Note that there is no correction at the $\p^3$ order, when written in terms
of conjugate momenta $p_i$, 
and it is tempting to conjecture that it would  
 be the case to all orders in $\p$. Yet they contain all order corrections
in $\p$ if written in terms of $v^i$. 
As the effective Lagrangian contains fermions, it has the second class 
constraints which can be analyzed using Dirac brackets.
Since it is non-local, the resultant
Dirac brackets are nontrivial, which is another source for non-trivial higher
order corrections.
The supercharges satisfy the usual supersymmetry algebra: 
\bea
[Q_a,Q_b]_+&=&{2\over p_-}\CH_{eff}\d_{ab} +\CO(\p^4)~,
\label{algebra}
\eea
where $\CH_{eff}$ is the effective Hamiltonian given by
\be
\CH_{eff}=p_-{p^2\over 1+\sqrt{1+hp^2}}-{ip_-\over 8}
{p^2p_i\rd_j h \p\g^{ij}\p \over\sqrt{1+hp^2}(1+\sqrt{1+hp^2})}
+\CO(\p^4)~.
\label{ham1}
\ee
\section{Supersymmetry of eleven-dimensional supergraviton in the light-cone
gauge}
In this section we consider the supersymmetry transformation rules of
 eleven-dimensional supergraviton on light-like compactified spacetime.
The natural gauge choice is the light-cone gauge and the supersymmetries which
preserve this gauge choice  
are identical to those of D0 quantum mechanics as expected. 

Consider the eleven-dimensional manifestly spacetime supersymmetric massless 
point particle action, 
\begin{equation}
S = \int d\lambda {\cal L } = {1 \over 2} \int 
d\lambda e^{-1}(\dot{x}^\mu+i\bar{\th}\Gamma^\mu\dot{\th})^2 ~,
\label{act1}
\end{equation}
where $\th$  are 32 component real 
spinors and $\bar{\theta}=\theta^T\Gamma^0$\footnote{
The eleven-dimensional $32 \times 32$ gamma matrices 
$\Gamma^r$ ($r = 0, 1, \cdots , 9, 11$)
that we use in this paper are given by
$$
\Gamma^0  = \left( \begin{array}{cc}
                            0 &   I_{16} \\
                    -I_{16}   & 0    \end{array} \right) ~ , ~
\Gamma^i  = \left( \begin{array}{cc}
                            0 &   \gamma^i \\
                    \gamma^i   & 0    \end{array} \right) ~ , ~
$$
$$ \Gamma^{11} = \Gamma^0 \cdots \Gamma^9 = 
 \left( \begin{array}{cc}
                            I_{16} &   0  \\
                   0  &  -I_{16}    \end{array} \right) ~ , ~
$$
where $I_{16}$ is the $16 \times 16$ identity matrix
and $\gamma^{i}$ are real $16 \times 16$ $SO(9)$ gamma 
matrices ($i=1 , \cdots, 9 $).
These gamma matrices satisfy $(\Gamma^0)^{\dagger}
= - \Gamma^0$ and $(\Gamma^i )^{\dagger} = \Gamma^i$.  
}.
This action describes eleven-dimensional supergraviton multiplet 
in flat Minkowskian spacetime.
The equations of motion of the multiplet are given by 
those of linearized eleven-dimensional supergravity.
The model has world-line reparametrization invariance under which 
$x^\mu$ and $\th$ transform as world-line scalar,
\bea
\d_\zeta x^\mu= \zeta {\dot x}^\m, \ \ \d_\zeta \th = \zeta {\dot \th}.
\label{diff}
\eea
 As we take the light-cone coordinate $x^-$ periodic, it is natural to 
identify $x^+$ as time coordinate. The natural gauge choice for the world-line
diffeomorphism in this DLCQ formulation is the static gauge 
\be 
\dot x^+=2\dot x^\t=2.
\label{fix1}
\ee
The target spacetime supersymmetry transformation laws with parameter $\xi$ 
are given by
\bea
\delta\th=\xi, &\ \ & \delta x^\mu = -i\bar{\xi}\Gamma^\mu\th~, \nn\\
\delta\bar{\th}=\bar{\xi}, &\ \ & \delta e =0 ~. 
\label{transa}
\eea
In addition, the action has local fermionic symmetry with parameter
$\k(\l)$, under which the fields transform as
\be
\d\th=ie\G\cdot p \k~,\ \ \d x^\mu = -i\bar{\th}\G^\mu\d\th~, \ \
\d e= 4e \dot{\bar{\th}}\k~,  
\label{transb}
\ee
where $p_\m=e^{-1}\eta_{\m\n}(\dot{x}^\n+i\bar{\th}\Gamma^\n\dot{\th})$
denotes the conjugate momentum of $x^\m$.
As being local gauge symmetry, this $\k$-symmetry 
reduces the $\th$ degrees of freedom by half.
We can fix it by choosing
\be
\G^+\th=0~, \ \ \th={\th_{(16)} \choose -\th_{(16)}}~, 
\label{fix2}
\ee
where $\G^{\pm}=(\G^{10}\pm \G^0)$.
Note that with this gauge 
fixing the conjugate $\bar{\theta} \equiv \theta^T \Gamma^0$
becomes
$$
\bar{\theta} =  \theta \frac{1}{2} ( \Gamma^+ -
\Gamma^- ) =  \theta \Gamma^{\tau} ~,
$$
where $\Gamma^{\tau} = \Gamma^+ / 2$.  This
is in accord with the gauge fixing (\ref{fix1}).

The light-cone momentum $p_-$, which is conjugate to the periodically 
identified light-cone coordinate $x^-\equiv x^-+2\pi R$, is quantized and 
given by $p_-={N\over R}$. Since the coordinate $x^-$ is cyclic, $p_-$ is 
conserved and we consider
the fixed $N$-sector of the theory. Therefore the appropriate effective action 
is given by Routhian
\begin{equation}
{\cal L}_{eff} = {\cal L } - p_- \dot{x}^-
 (p_- ) = -p_-\dot{x}^-~,
\label{routh}
\end{equation}
where the last relation in the above comes from the constraint equation
\begin{equation}
 \eta_{\m\n} p^\m p^\n =0~,
\label{const}
\end{equation}
which is also used to solve $\dot{x}^-$ in terms of $p_-$, $x^i$, $\th$.  

The effective Lagrangian after the gauge fixing (\ref{fix1}) and (\ref{fix2})
becomes
\be
\CL_{eff}=p_-({(v^i)^2\over 2}+4i\th_{(16)}\dot{\th}_{(16)})~,
\label{act2}
\ee
which corresponds to the (0+1)-dimensional $U(1)$ supersymmetric 
Yang-Mills quantum mechanics with
adjoint fermions $\psi\equiv2\sqrt{2}\th_{(16)}$. 

This tells that 
even though the original supersymmetry transformation laws of 
the first-quantized supergravity action (\ref{act1}) are very different
from those of super Yang-Mills theory, they should be the same after 
the gauge fixing (\ref{fix1}) and (\ref{fix2}).
Half of the original supersymmetries with $\Gamma^+ \xi=0$ remain 
to be true symmetry in which 
$$
\d\th=\xi~, \ \ \ \d x^i=0~,
$$
even after gauge fixing.
These are just constant shift in spinors which also appear as trivial 
kinematic
supersymmetry in the D0 quantum mechanics.

On the other hand, other half of the form 
$\xi={\xi_{(16)}\choose \xi_{(16)}}$, 
corresponding to $\G^-\xi= 0$, do not preserve the gauge fixing 
conditions and thus
the original supersymmetry transformation laws (\ref{transa}) with these
parameters should 
be modified in such a way to preserve those by taking 
appropriate $\kappa$-transformations (\ref{transb}). From the condition 
that total transformations should preserve the gauge fixing (\ref{fix2}):
\be
\G^+(\d\th)=\G^+(\xi+i\G\cdot p\k)=0~,
\label{transc}
\ee
we find the relation between target spacetime 
supersymmetry parameter $\xi$ and
the kappa symmetry parameter $\k$ of the form, 
\be
\k={\k_{(16)} \choose -\k_{(16)}}~,
\label{kappa}
\ee 
which is given by
\[ 
\k_{(16)} = {i\over 2}\xi_{(16)}~.
\]
The total supersymmetry transformations preserve the static gauge (\ref{fix1}) 
and are given by
\bea
\d\psi=v^i\g^i\e~, \nn \\
\d x^i=-i\e\g^i\psi
\label{transd}
\eea
with $\e\equiv\sqrt{2}\xi_{(16)}$.
These are nothing but the supersymmetry transformation laws of 
(0+1)-dimensional super Yang-Mills quantum mechanics
at the tree level. 

This suggests that the matrix model, viewed as the large $N$ limit of DLCQ
M theory, in the infrared limit may describe eleven-dimensional 
second-quantized DLCQ supergravity, i.e. the large $N$ limit of 
first-quantized DLCQ supergravity. To confirm this conjecture, one may
construct vertex operators \cite{taylor,green} 
which describe generic interactions among various fields and 
macroscopic objects and compare with super Yang-Mills theory, 
which is outside the scope of this paper. 

In the next section we consider the supergraviton in gravitational wave
background, which would correspond to the case 
of two-body interactions between two clusters of D0-particles in which one 
plays the role as a source and the 
other as a probe. 

\section{Supergraviton in gravitational wave background}
The action (\ref{act1}) can be generalized to describe supergraviton  
on the general background in the following form:     
\begin{equation}
S = \int d\lambda {\cal L } = {1\over 2}\int  
d\lambda e^{-1} \eta_{rs} \Pi^r \Pi^s ~,
\label{act3}
\end{equation}
where $ Z^M (\lambda ) = (x^{\mu} (\lambda) ~ , 
\theta^{a} (\lambda) ) $ are the superspace 
coordinates.\footnote{In the superspace formalism, 
the indices $(A, B, C, \cdots)$ 
collectively denote the bosonic and fermionic tangent space 
indices, while  
$(M, N, P, \cdots)$  collectively denote the 
bosonic and fermionic curved space indices. 
In its component form, we use 
$(\mu, \nu, \rho, \cdots )$ 
indices for the curved space bosonic
coordinates, and $(r, s, t, \cdots )$ indices for the tangent space bosonic
coordinates.
Therefore the metric $\eta_{rs}$ is the Lorentz invariant
constant metric.  
Spinor indices are denoted as $(a, b, c, \cdots)$.}
 The pull-back $\Pi^A$ of the supervielbein 
$E_{M}^{A}$ to the particle's
world-line satisfies $\Pi^A = ( d Z^M / d \lambda ) E_M^A$. 
This can be determined order by order in $\th$ 
by following the method in \cite{dpp} and is given by
\begin{equation}
 \Pi^r = \dot{x}^{\mu} (e_{\mu}^r 
-\frac{1}{4} \bar{\theta} \Gamma^{rst} \theta
\omega_{\mu st } )
+ \bar{\theta} \Gamma^r \dot{\theta} + \CO(\th^4) 
\label{a1}
\end{equation}
for the background geometry with vanishing
three form gauge field $C_{\m\n\rho}$ and gravitino $\p_\m$.

\subsection{Effective action}
As the background geometry we consider is independent of periodically 
identified coordinate $x^-$, the quantized conjugate momentum is still
conserved and, 
in the fixed $p_-$ sector, the effective action is again of the form 
(\ref{routh}). 
 In \cite{hks3}, along the same steps as described above in the free massless 
superparticle case, it was shown that the supersymmetric effective 
action, in the light-cone gauge, corresponding to $v^4$ term 
is the same as  matrix theory effective action. The action has world-line
diffeomorphism and $\kappa$-symmetry as local gauge symmetries.
We use the same gauge fixing conditions (\ref{fix1}) and (\ref{fix2}) as in 
the case of flat spacetime background.

The eleven-dimensionally lifted D0-solution with asymptotically light-like
compactification, $x^-\equiv x^-+2\pi R$, is given by \cite{hks}
\begin{equation}
ds_{11}^2 = dx^+ dx^- + h(r) (dx^-)^2 + dx_1^2 + \cdots
   + dx_9^2 ~,
\label{11met1}
\end{equation}
where $r=(x_1^2+\cdots x_9^2)^{1/2}$ and $\psi_\m=C_{\m\n\r}=0$.
The harmonic function $h(r)$, which characterize this background geometry, 
is of the form 
$$ 
h(r) = {15 \over 2} 
  \sum_{I=1}^N \frac{l_p^9/R^2}{| \vec{r} - \vec{r_I} |^7}~,
$$
where $l_p$ is eleven-dimensional Planck scale. 
This geometry represents half BPS states of $M$ theory, admitting 
16 Killing spinors which
satisfy the Killing spinor equation
\begin{equation}
 D_{\mu} \xi = (\partial_\mu
- \frac{1}{4} \omega_{\mu}^{rs} \Gamma_{rs} ) \xi = 0 ~.
\label{killing}
\end{equation}
These sixteen Killing spinors are of the form
\bea
 \xi =\left( \begin{array}{c}
                         \xi_{(16)}  \\
          \xi_{(16)}     \end{array} \right)= {1\over \sqrt{2}}f^{-1/4}
\left( \begin{array}{c}
                         \e  \\
          \e     \end{array} \right)~,
\label{kills}
\eea   
where $\e$ denotes a constant spinor\footnote{From now on all spinors 
 denote 16-component spinors and subscripts are omitted.} 
and $f(r)=1+h(r)$.

The effective Lagrangian (\ref{routh}) on this background can be decided order
by order in $\p$\footnote{
$\psi_a$ is defined by $\psi_a\equiv 2\sqrt{2}f^{-1/4}\th_a$. These play the 
role as the superpartner of Higgs field $x^i$ in the context of 
Yang-Mills quantum mechanics and thus the supersymmetry
transformation laws for these turn out to be much simpler than those of $\th$. 
In deriving various formula, we will switch back and forth
between $\psi_a$ and $\th_a$, but the final results are presented in terms of 
$\psi_a$.} using the constraint equation,
\[
\eta_{rs} \Pi^r \Pi^s = 0~.
\]
It is given by, up to $\psi^4$ order,  
\bea
{1\over p_-}\CL_{eff}={v^2\over 1+\sqrt{1-hv^2}}
           +i{(1+\sqrt{1-hv^2})\over 4\sqrt{1-hv^2}}
           \psi\dot{\psi}+i{p_-v^2 v_i\partial_j h(\psi\g^{ij}\psi)
           \over 8(1-hv^2+\sqrt{1-hv^2})}
          +\CO(\psi^4)~.
\label{eff1}
\eea
In contrast to \cite{hks3}, in which only $v^4$ terms and their superpartners 
are presented, we retain the expression to all orders in $v^i$. 
This may correspond to the effective Lagrangian of a D0-brane probe 
in the background of N D0-branes, relatively moving with the velocity $v^i$
in the transverse direction.
There is a non-renormalization theorem for the $v^4$ term and 
its superpartners in the (0+1)-dimensional super Yang-Mills quantum mechanics
\cite{sethi2,lowe}.
Therefore they are completely determined by the one-loop
calculations and was shown to agree with the 
supergravity result (\ref{eff1}) \cite{hks3}.
The same arguments hold true for the $v^6$ term and 
its superpartners and they agree 
with the supergravity result (\ref{eff1}) as well \cite{sethi3}.

\subsection{Supercharge: leading order corrections}
In the matrix model, 
it is extremely tedious to determine the supersymmetry 
transformation laws, though still   
one can decide the effective
action of bosonic $v^4$ term and its superpartners without much knowledge on 
those \cite{sethi2,hks4}. 
In the supergravity side, the explicit form of supersymmetry transformation 
rules are known up to $\psi^3$ order. 
We use these to find supercharges for supergraviton in the 
gravitational wave background. Up to $v^6$ (and possibly to all order in $v$)
they should correspond to the supercharges in the matrix theory. 

The supersymmetry transformation laws of fields in the supergraviton 
action (\ref{act3}), up to $\th^3$ order, are given by 
\bea
\d\th &=&\xi+
{i \over 4}(\bar{\th} \G^{\mu}\xi)\o_\m^{rs}\G_{rs}\th +\CO(\th^4)~,\nn\\ 
\d x^\mu &=& i\bar{\th}\G^\mu\xi +\CO(\th^3)~.
\label{transe}
\eea
As mentioned earlier,
the action (\ref{act3}) has local fermionic $\k$-symmetry, whose   
transformation laws are 
\bea
\d\th&=&i\G\cdot \Pi \k +
{1 \over 4}(\bar{\th} \G^{\mu}\G\cdot\Pi\k)\o_\m^{rs}\G_{rs}\th 
+\CO(\th^4)~,\nn\\ 
\d x^\mu& =& -i\bar{\th}\G^\mu\d\th~ +\CO(\th^3),\label{transf} \\
\d e&=& 4e \dot{\bar{\th}}\k~ +\CO(\th^3)~.\nn  
\eea

Obviously the sixteen Killing spinors (\ref{kills}), which 
correspond to the dynamical 
supersymmetry in the flat background limit, do not respect the gauge
fixing condition (\ref{fix2}). 
At the leading order in $\th$, to preserve the gauge fixing condition 
(\ref{fix2}), the transformation should be supplemented 
by the 
$\k$-transformations of the form (\ref{kappa}) with
\bea 
\k= i{f^{1/2}\over 1+ \sqrt{1-hv^2}}\xi 
+\CO(\th^2)~.
\label{kappa2}
\eea
The combined transformation law of (\ref{kills}) and (\ref{kappa2}) 
for $\th$ is given by 
\bea
\d\th ={f^{1/2}\over 1+\sqrt{1-hv^2}}(v\cdot \g)\xi 
+\CO(\th^2)~.
\label{trans4a}    
\eea

In contrast to the flat background case, the combined transformation laws 
 do not satisfy the 
gauge fixing (\ref{fix1}) as $\d x^+$ is non-vanishing: 
\bea
\d x^+ = 4i{h\over (1+\sqrt{1-hv^2})}(\th v\cdot \g\xi) +\CO(\th^3)~.     
\label{trans4b}
\eea
This should be compensated by world-line reparametrization with parameter
\be
\zeta=-2i{h\over (1+\sqrt{1-hv^2})}(\th v\cdot \g\xi) +\CO(\th^3)
\label{diff1}
\ee
in (\ref{diff}).
This in turn gives additional transformations on $x^i$ and $\th_a$.
The overall transformation law for $x^i$, in the leading order in $\th$, 
is given by
\bea 
\d x^i&=& 4i(\th\g^i\xi) +\zeta{\dot x}^i +\CO(\th^3)\nn \\
      &=& 4i(\th\g^i\xi) -2i{h(\th v\cdot\g\xi)\over (1+\sqrt{1-hv^2})} 
         {\dot x}^i +\CO(\th^3).
\label{trans4c}
\eea
The effective Lagrangian (\ref{eff1}) is indeed invariant under the 
transformation laws (\ref{trans4a}) and (\ref{trans4b}), in the 
leading order terms of $\psi$, up to total derivatives, 
\[
\d \CL_{eff}= -i{p_-\over 2}{d\over d\l} (\e v\cdot\g \p) +\CO(\p^3)~. 
\]

The corresponding supercharges can be easily obtained by Noether method 
and are given by the simple form:
\bea
Q= p\cdot\g\p +\CO(\p^3).
\label{charge2}
\eea

\subsection{Dirac brackets and supersymmetry algebra}
In order to see the above supercharges (\ref{charge2}) give the right 
transformation laws (\ref{trans4a}) and (\ref{trans4b}), 
we need to study carefully the commutation 
relations among fields.   From the effective Lagrangian (\ref{eff1})
 one can read off the effective conjugate momenta $p_i$ 
of $x^i$ and $\pi_a$ of $\psi_a$: 
\bea
p_i\equiv{1\over p_-}{\rd \CL_{eff}\over \rd v^i}
&=&{v_i\over \sqrt{1-hv^2}} +{ih(\p\dot{\p})v_i\over 4(1-hv^2)^{3/2}}
+{iv_k\rd_jh(\p^T\g^{kj}\p)v_i\over 8(1-hv^2)^{3/2}}\nn \\ 
&&+{iv^2\rd_jh(\p^T\g^{ij}\p)\over 8(1+\sqrt{1-hv^2})\sqrt{1-hv^2}}
+\CO(\psi^4)~, 
\label{momenta1}
\eea
and
\bea
\pi_a\equiv{1\over p_-}{\rd_r \CL_{eff}\over \rd \dot{\psi_a}}
={i(1+\sqrt{1-hv^2})\p_a\over 4\sqrt{1-hv^2}}+\CO(\psi^3)~,
\label{momenta2}
\eea
which has been defined in units of $p_-$ for simplicity.
Note that (\ref{momenta2}) gives the second-class constraints:
\bea
\Phi_a=\pi_a-{i\over 4}(1+\sqrt{1+hp^2})\psi_a+\CO(\psi^3)\approx 0
\label{const1}
\eea 

It is well-known that in order to deal with the second-class constraints 
in the 
Hamiltonian formalism, which
is typically present in the theory involving fermions, we need to introduce 
Dirac brackets. Let's denote $(q^A, p_A)$ as fields and their conjugates
collectively for bosons and fermions.
Then Poisson bracket between two functions $\CF(q,p)$ and $\CG(q,p)$ 
is defined as 
\bea
\{ \CF, \CG\}_{PB}\equiv {\rd_r \CF\over\rd q^A}{\rd_l \CG\over\rd p_A} 
-(-1)^{n_\CF n_\CG}{\rd_r \CG\over\rd q^A}{\rd_l \CF\over\rd p_A}~, 
\eea
where $n_\CF$ is the fermion number of $\CF$.
If the theory contains the 
second class constraints, $\Phi_a (q^A, p_A) \approx 0$, the Poisson bracket 
should be replaced by the
Dirac bracket which is defined as
\bea
\{ \CF, \CG\}\equiv \{\CF,\CG \}_{PB}
-\{ \CF, \Phi_a\}_{PB}(\CA^{-1})_{ab} \{\Phi_b,\CG \}_{PB},  
\eea
where $\CA_{ab}\equiv\{ \Phi_a, \Phi_b\}_{PB}$.

In the case we consider, the Poisson brackets among the fields and their 
conjugates are
given by
\bea
\{x^i, p_j\}_{PB}=\d^i_j~, \ \ \ \{\psi_a, \pi_b\}_{PB}=\d_{ab}
\eea
and all others vanish.
For the constraint (\ref{const1}), $\CA_{ab}$ read
\bea
\CA_{ab}
=-{i\over 2}(1+\sqrt{1+hp^2})\d_{ab}+\CO(\psi^2)~,
\label{pb1}
\eea
from which the Dirac brackets among fields and their conjugates
can be read as
\bea
\{x^i, p_j\}&=&\d^i_j~ +\CO(\psi^2)~, \nn\\
\{\psi_a, \pi_b\}&=&{1\over 2}\d_{ab}+\CO(\psi^2)~, \nn\\
\{\psi_a, \psi_b\}&=&-{2i\over (1+\sqrt{1+hp^2})}\d_{ab}+\CO(\psi^2)~,
\label{db1}\\
\{x^i, \psi_a\}&=&-{hp_i\over 2\sqrt{1+hp^2}(1+\sqrt{1+hp^2})}
\psi_a+\CO(\psi^3)~, \nn\\
\{p_i, \psi_a\}&=&{p^2\rd_ih\over 4\sqrt{1+hp^2}(1+\sqrt{1+hp^2})}
\psi_a+\CO(\psi^3)~.\nn
\eea
Note that the Dirac brackets between bosonic and fermionic variables
 are highly nontrivial
representing the nonlocal nature of the effective theory. 

Indeed using these Dirac brackets (\ref{db1}) one finds the transformation 
laws:
\bea
\d x^i&=&-i\e_a\{Q_a,x^i\}= i\p\g^i\e
   -i{hv^i (\p v\cdot\g\e) \over 2(1+\sqrt{1-hv^2})}
     +\CO(\p^3), \nn \\ 
\d\p_a&=&-i\e_b\{Q_b,\p_a\}={2\over(1+\sqrt{1-hv^2})}(v\cdot\g\e)_a
+\CO(\p^2)~,
\eea 
which agree with (\ref{trans4a}) and (\ref{trans4c}) for 
$\psi= 2\sqrt{2}f^{-1/4}\th$.

The supersymmetry algebra can be also read  
 using these Dirac brackets as
\bea
\{Q_a,Q_b\}&=&-i{2\over p_-}\CH_{eff}\d_{ab} +\CO(\p^2)~.
\eea
By replacing the Dirac bracket with the (anti-)commutator,
$[Q_a,Q_b]_+=i\{Q_a, Q_b\}$, we get (\ref{algebra}) up to $\p^2$ order. 

\subsection{Higher order corrections}
In this section we find the next-to-leading order corrections to the 
supercharge from the transformation laws (\ref{transe}).
After some calculations, we find the supersymmetry should be modified by
$\k$-transformation (\ref{kappa}) $\k^{(2)}$ at the $\p^2$ order of the form: 
\bea 
\k^{(2)}&=& 
{2(\th v\cdot\g\xi)\over (1+ \sqrt{1-hv^2})^2}\partial_i h\g^i\th 
-{hv^2v_i\partial_j h(\th\g^{ij}\th)\over (1+\sqrt{1-hv^2})^3\sqrt{1-hv^2}}
\xi \nn \\
&&-{2h(\th\dot{\th})\over (1+\sqrt{1-hv^2}) \sqrt{1-hv^2}}\xi~.
\label{kappa3}
\eea
The resultant transformation law for $\th$ at the next-to-leading order, 
which includes the transformation 
due to the above $\k$-transformation (\ref{kappa3}) as well as world-line 
diffeomorphism
by $\zeta$ (\ref{diff1}), becomes  
\bea
\d^{(2)}\th &=& if^{-1}\partial_i h(\th\g^i \xi)\th 
+i{hv^2v_i\rd_j h(\th\g^{ij}\th) (v\cdot \g)\xi 
\over(1+\sqrt{1-hv^2})^3\sqrt{1-hv^2}}
+i{2h(\th\dot{\th})(v\cdot \g)\xi\over (1+\sqrt{1-hv^2})\sqrt{1-hv^2}}\nn \\
&&+i{v^2\rd_j h(\th\g^{ij}\th)\g^i\x - 2(\th v\cdot\g\xi)v^i\rd_j h
\g^i\g^j\th \over(1+\sqrt{1-hv^2})^2}   
-i{2h(\th v\cdot\g\xi)\over (1+\sqrt{1-hv^2})}\dot{\th}~.\nn    
\eea
This becomes much simpler if we impose the effective equations of motion:
\bea
\dot{\p}&=&-{v^2\over 2(1+\sqrt{1-hv^2})^2}v_i\rd_j h \g^{ij}\p 
   -{{d\over dt}(hv^2)\over 4(1+\sqrt{1-hv^2})(1-hv^2)}\p +\CO(\p^3)~,\nn\\
{d\over dt}(hv^2)&=&{2(1-hv^2)v^2\over 1+\sqrt{1-hv^2}}{dh\over dt}
+\CO(\p^2)~, 
\label{eom}
\eea 
which may be considered as the
full quantum corrected equations of motion of the matrix quantum mechanics.
Up to these equations of motion, the on-shell 
transformation laws for Yang-Mills fermions $\psi$, at the $\p^2$ order, 
becomes
\be
\d^{(2)}\p=i{ v^2\rd_j h(\p\g^{ij}\p)\over 4(1+\sqrt{1-hv^2})^2}\g^i\e
+i{(\e v\cdot\g\p)\over 4(1+\sqrt{1-hv^2})}(v_i\rd_j h\g^{ij}+{dh\over dt})\p~.
\label{trans4e}
\ee

Now we would like to determine the next-to-leading order corrections to
the supercharges which gives the above transformation law (\ref{trans4e}).
As a commutation relation with supercharge, the  $\p^2$ order corrections 
in (\ref{trans4e}) come from the $\p^3$ order corrections of 
the supercharge and/or the constraints (\ref{const1}) which give rise to 
the higher order corrections in the Dirac brackets. 

Let the 
conjugate momenta of the $\p$'s are of the form   
\bea
\pi_a={i\over 4}(1+\sqrt{1+hp^2})\psi_a
      +l_{abcd}\p_b\p_c\p_d+\CO(\p^5)~,
\label{mom3}
\eea
where $l_{abcd}=l_{a[bcd]}(x^i,p_i)$
are totally antisymmetric in last three indices.  
We assumed that $\pi_a$ do not depend on $\dot{\p}$ when written in terms of
phase space variables $x^i$, $p_i$. If not, these do not give the 
constraints, invalidating the analysis in the lower order in $\psi$. This is
justified by the consistency of the results followed. Of course, when  
rewritten in terms of $x^i$ and $v^i$, they have  $\dot{\p}$ dependence.
It will be interesting to justify these by obtaining the higher order 
corrections in $\p$ to the effective action from direct calculations 
following the method outlined in \cite{dpp}.     

The relations (\ref{mom3}) give $\p^3$ order corrections to the second class 
constraints 
and the Poisson brackets between the constraints become 
\bea
\CA_{ab}
=-{i\over 2}(1+\sqrt{1+hp^2})\d_{ab}-3(l_{abcd}+l_{bacd})\p_c\p_d+\CO(\psi^4).
\eea
The Dirac brackets between the spinors $\p_a$, which include the
$\p^2$ order corrections, are given by
\bea
\{\psi_a, \psi_b\}=-{2i\over (1+\sqrt{1+hp^2})}\d_{ab}
+{12(l_{abcd}+l_{bacd})\over (1+\sqrt{1+hp^2})^2}\p_c\p_d+\CO(\psi^4)~.
\label{db2}
\eea
The supercharge may also have the $\p^3$ order corrections 
and thus generically
can be written as
\bea
Q_a= p\cdot\g_{ab}\p_b +A_{abcd}\p_b\p_c\p_d+\CO(\p^5)~,
\label{charge3}
\eea
where  $A_{abcd}=A_{a[bcd]}(x^i,p_i)$ are also totally antisymmetric in last
three indices.
Note that the momenta $p_i$ in (\ref{mom3}) and (\ref{charge3}) have $\p^2$
order corrections when written in terms of $v^i$. 

In order to decide $A_{abcd}$ and $l_{abcd}$, we 
calculate $\{Q_a,\p_b\}$ and $\{Q_a,Q_b\}$ and compare with
(\ref{trans4e}) and (\ref{algebra}). 
These consistency conditions give the unique 
choice for $A_{abcd}$ and $l_{abcd}$. 
The transformation laws for $\p_a$  with
the supercharge (\ref{charge3}) can be read readily using (\ref{db1}) and 
(\ref{db2}) and are given by
\bea 
\{Q_a, \psi_b\}
   &=&-i{2p\cdot\g_{ab}\over (1+\sqrt{1+hp^2})}
      +{p^2\rd_j h\g^j_{ac}\p_c\over
       4\sqrt{1+hp^2}(1+\sqrt{1+hp^2})}\p_b \\
   &&+6{[(1+\sqrt{1+hp^2})A_{abcd}+2p\cdot\g_{ae}
     (l_{becd}+l_{ebcd})] \over (1+\sqrt{1+hp^2})^2}
      \p_c\p_d+\CO(\psi^4)~.\nn
\eea
By comparing with (\ref{trans4e}), we find a relation:
\bea
&& 24\sqrt{1-hv^2}\biggl[A_{abcd}+{2v\cdot\g_{ae}\over (1+\sqrt{1-hv^2})}
(l_{becd}+l_{ebcd})\biggr]\p_c\p_d 
\label{cond1}\\
&=&\biggl[v\cdot\g_{ac}(v_i\rd_j h\g^{ij}_{bd}+{dh\over dt}\d_{bd})
-v\cdot\g_{ab}v_i\rd_j h\g^{ij}_{cd}-v^2\rd_j h\g^j_{ac}\d_{bd}\biggr]
\p_c\p_d~.\nn
\eea
The commutation relations among $Q_a$'s read
\bea 
\{Q_a, Q_b\}&=&-2i{p^2\d_{ab}\over (1+\sqrt{1+hp^2})}
      -{p^2\over 4}{[p\cdot\g_{ac}\rd_j h\g^j_{bd}+(a\leftrightarrow b)]\over
       \sqrt{1+hp^2}(1+\sqrt{1+hp^2})}\p_c\p_d \nn\\
      &&+6{p\cdot\g_{be}\biggl[(1+\sqrt{1+hp^2})A_{aecd}
      +p\cdot\g_{af}
     (l_{efcd}+l_{fecd})\biggr]+(a\leftrightarrow b) 
     \over (1+\sqrt{1+hp^2})^2}\p_c\p_d
      +\CO(\psi^4)~.\nn
\eea 
These satisfy (\ref{algebra}) provided that 
\bea
&&\biggl[v\cdot\g_{be}\biggl(A_{aecd}+{v\cdot\g_{af}\over (1+\sqrt{1-hv^2})}
 (l_{efcd}+l_{fecd})\biggr)+(a\leftrightarrow b) \biggr]\p_c\p_d\nn\\
=&&{v^2\over 24\sqrt{1-hv^2}}\biggl[\biggl(v\cdot\g_{ac}\rd_j h\g^j_{bd}
+(a\leftrightarrow b)\biggr)-v_i\rd_j h \g^{ij}_{cd}\d_{ab}\biggr]\p_c\p_d.
\label{cond2}
\eea
From (\ref{cond1}) and (\ref{cond2}), we find the unique solution
\bea
A_{abcd}&=&0~, \nn\\ 
l_{abcd}&=&-{1+\sqrt{1+hp^2}\over 32\sqrt{1+hp^2}}p_i\rd_j h\d_{a[b}\g_{cd]}~,
\eea
and therefore the supercharges are given by (\ref{charge1}).

From this one can read off the $\p^3$ order corrections to the 
supersymmetry transformation laws of $x^i$. 
The Dirac brackets between $x^i$ and
$\p_a$ are given by 
\bea
\{x^i, \psi_a\}&=&-{hp_i\over 2\sqrt{1+hp^2}(1+\sqrt{1+hp^2})}
    \psi_a 
    -i{\rd_j h(\p\g^{ij}\p)\over \sqrt{1+hp^2}}\p_a \nn\\
     && +i{(2+\sqrt{1+hp^2})hp_i (p_k\rd_j h\p\g^{kj}\p)
       \over 16(1+\sqrt{1+hp^2}) (1+hp^2)^{3/2}}\p_a
      +\CO(\psi^5)~,
\label{db3}
\eea
and therefore $x^i$ transform as
\bea
\d x^i&=& i\p\g^i\e-i{hv^i (\p v\cdot\g\e) \over 2(1+\sqrt{1-hv^2})}
     -{h(v_k\rd_j h\p\g^{kj}\p)(\p v\cdot\g\e)\over 8(1+\sqrt{1-hv^2})^2}v^i\\
     && +{(\rd_j h\p\g^{ij}\p)(\p v\cdot\g\e)\over 8(1+\sqrt{1-hv^2})}
      + {hv^2\rd_j h(\p\g^{kj}\p)(\p \g^k\e)
      \over 16(1+\sqrt{1-hv^2})^2}v^i +\CO(\p^5)~.\nn  
\eea 

\section{Discussions}
In this paper we found an explicit form of the supercharges, up to
the $\p^5$ order, of the
supergraviton in the background of eleven-dimensionally lifted D0 geometry.
They should correspond to
the supercharges of the effective action of (0+1)-dimensional
matrix quantum 
mechanics for, at least, $v^4$ and $v^6$ order terms and their 
superpartners. From the perspectives of matrix quantum mechanics, the simple 
form of the supercharges (\ref{charge1}) is quite striking. 
Note also that this simple form makes it an easy step to go from classical, 
represented by Dirac brackets, to quantum, represented by commutators,
without operator ordering ambiguity. 
It would be very
interesting to see whether it holds true to all orders in $\p$.
We expect similar results hold for higher-dimensional Yang-Mills theories with
16 supercharges which correspond to DLCQ M theory on torus 
compactification. These are under investigations.

\begin{center}
{\bf Acknowledgements}
\end{center}

I would like to thank to Jin-Ho Cho, Youngjai Kiem,
Julian Lee, Sangmin Lee, Jae-Suk Park and Hyeonjoon Shin 
for useful discussions.

\end{document}